\documentclass[12pt]{iopart}
\usepackage{graphicx}
\usepackage{cite}
\newcommand{\dzup}[3]{\frac{d^{#3}#1}{d#2^{#3}}}

\newcommand{\ctg}{\mbox{ctg}}
\newcommand{\sgn}{\mbox{sgn}}
\newcommand{\sab}{\sigma_{lv}}
\begin{document}
\jl{3}
\title{Liquid drop in a cone - line tension effects}
\author{G P Kubalski and M Napi\'{o}rkowski}
\address{Institute of Theoretical Physics, Warsaw University, 
Ho\.za 69, 00-681 Warsaw, Poland}

\begin{abstract}
The shape of a liquid drop placed in a cone is analyzed macroscopically.
Depending on the values of the cone opening angle, the Young angle 
and the line tension four different interfacial configurations may be  
realized. The phase diagram in these variables is constructed and discussed; 
it contains both the first- and the second-order transition lines.  
In particular, the tricritical point is found and the value of the critical 
exponent characterizing the behaviour of the system along the line of 
the first-order transitions in the neighbourhood of this point is 
determined. \\
\end{abstract}
\pacs{68.55.Jk, 68.35.Rh, 68.45.Gd}
\maketitle

\section{Introduction}

The role of line and surface tensions in determining the morphology 
of interfaces is well known \cite{W,dG,D,RW,LaI,Indekeu,BaN}. For 
example, the values of the contact angles formed by an interface and a 
substrate depend on both the line and the surface tensions. 
In the simplest case of a drop sessile on a flat substrate 
the influence of line tension is described by the modified Young equation 
\cite{RW,BaN}. Under special circumstances one can even expect a competition 
between the interfacial tension and the line tension effects. This 
competition originates from the fact that while the interfacial tension 
is positive \cite{W10} and tends to decrease the interfacial area the 
negative values of the line tension \cite{W11} support the increase of the 
three-phase contact line. This competition between the line and the surface 
tension  effects may result in forming different interfacial configurations 
and transitions between them. For example, Widom \cite{W12} in his 
analysis of a sessile drop on flat substrate found a line of first-order 
phase transitions parameterized by the line tension. 

In this paper we analyze the equilibrium shapes of a non-volatile liquid 
drop placed in a cone by taking into account both the surface tension and 
the line tension effects. The system and our methods of macroscopic analysis 
are described in Section 2. In Section 3 we find the equilibrium 
configurations of the liquid which are parameterized by the values of the cone 
opening angle, the Young angle - which itself depends on the values of the 
interfacial and surface tensions - and the line tension. This section is 
divided into two parts. 
In the first part the case of zero line tension is considered analytically; 
in the second part we solve numerically the non-zero line tension case. The 
phase diagram found in this section is our main result. The last section 
contains the conclusions and remarks. 

\section{Macroscopic analysis}

\begin{figure}
\begin{center}
 \includegraphics{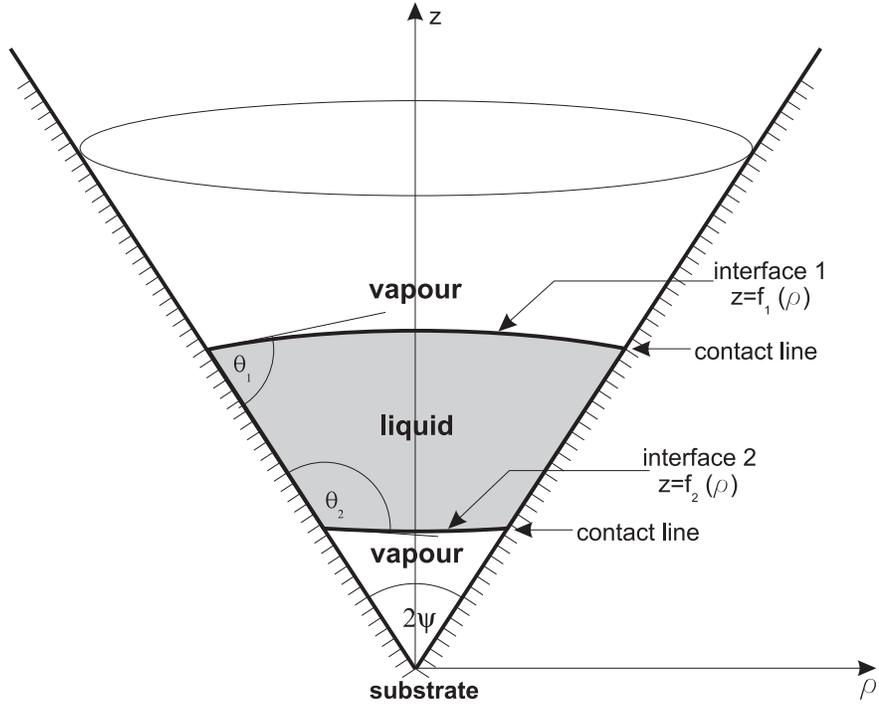}
\end{center}
\caption{The schematic plot of the system. As an  example the configuration 
with two interfaces is shown. The z-axis coincides with the axis of the
cone.\label{fsystem}}
\end{figure}

We consider a given amount of liquid placed on a nondeformable 
solid substrate which has the shape of a cone with the opening angle 
$2\psi$, $\psi < \pi/2$. The bulk thermodynamic conditions, say the 
temperature and the pressure are chosen such that the liquid is in 
equilibrium with its vapour. The liquid can form one or more separate 
droplets, see \fref{fsystem}. The physical parameters used in our analysis 
are the liquid-vapour, substrate-liquid, and substrate-vapour surface 
tensions 
denoted as $\sigma_{lv}$, $\sigma_{sl}$, $\sigma_{sv}$, respectively; 
the line tension accompanying the substrate-liquid-vapour contact line is 
denoted as $\eta_{slv}$. The gravity effects are neglected; for the 
discussion of their role see \cite{W12}. 

The configurations of the liquid are uniquely determined by the shapes 
of the liquid-vapour interfaces. The shape of the $i$-th interface is 
described by the function $f_{i}$. Thus the set $\{f_{i}\}_{i=1}^{n}$ 
describes the liquid configuration with  $n$ liquid-vapour interfaces 
numbered from the top to the bottom of the cone. Our analysis is 
restricted to configurations with cylindrical symmetry and interfaces which 
are smooth at the cone axis. In consequence all functions $f_{i}$ depend 
on only one variable - the distance $\rho$ from the cone axis.\\

We start by constructing the macroscopic constrained free energy which is 
a functional of interfacial configurations.  Then we find the equilibrium 
configurations which correspond to the minima of this functional. Actually 
we consider the excess free energy with respect to configuration in which 
the cone is filled the the vapour only. This excess free energy is denoted 
by ${\cal F}[\{f_{i}\}_{i=1}^{n}]$ and has the following form 
\begin{equation}
{\cal F}=A_{lv}\sigma_{lv} + A_{sl} (\sigma_{sl} - \sigma_{sv}) + L \eta_{slv} 
\quad ,
\end{equation}
where $A_{lv}$, $A_{sl}$ denote the area of the 
liquid-vapour and substrate-liquid interfaces, respectively and $L$ is the 
length of the three-phase contact line. After substituting the expression 
for $A_{lv}$, $A_{sl}$, and $L$ in terms of $\{f_{i}\}_{i=1}^{n}$ into 
equation (1) one obtains  
\begin{eqnarray}
\label{deffF}
\fl {\cal F}[\{f_{i}\}_{i=1}^{n}]= 2\pi \sab \int_{0}^{\infty}
  \left(\sum_{i=1}^{n} \theta(f_{i}(\rho) - \rho\ctg\psi) \left(\sqrt{1 + 
   \left(\dzup{f_{i}}{\rho}{}\right)^{2}}\right.\right. + \nonumber\\
 \left.\left.\frac{1}{\sin\psi} \left((-1)^{i} \frac{\sigma_{sv} - 
  \sigma_{sl}}{\sigma_{lv}} 
  + \frac{\eta_{slv}
  \sin\psi}{\sab\rho}\right)\right)\right) \rho d\rho \quad .
\end{eqnarray}
$\theta$ denotes the Heaviside function. We look for the minima of the above 
functional under the constraint of fixed volume $V$ of the liquid. Thus the 
functional to be minimized takes the form 
\begin{equation}
 {\cal F}^{*}[\{f_{i}\}_{i=1}^{n}]={\cal F}[\{f_{i}\}_{i=1}^{n}] + \lambda 
  V[\{f_{i}\}_{i=1}^{n}] \quad ,
\end{equation}
where $\lambda$ is the Lagrange multiplier and 
\begin{equation}
\label{warobj1}
 V = 2 \pi \int_{0}^{\infty} \left(\sum_{i=1}^{n} (-1)^{i+1} 
  (\theta(f_{i}(\rho) - 
  \rho\ctg\psi) (f_{i}(\rho) - \rho \ctg\psi)\right) \rho d\rho \quad .
\end{equation}
Accordingly  the equilibrium shape of the $i$-th liquid-vapour interface 
fulfills the following differential equation
\begin{equation}
\label{rownobj}
 \left(1 + \left(\dzup{f_{i}}{\rho}{}\right)^{2}\right)^{-3/2} 
  \left(\dzup{f_{i}}{\rho}{2} \rho +
  \left(1 + \left(\dzup{f_{i}}{\rho}{}\right)^{2}\right)
  \dzup{f_{i}}{\rho}{}\right) = (-1)^{i+1} \frac{\lambda}{
  \sab} \rho 
\end{equation}
supplemented by the boundary conditions

\begin{equation}
\label{defri}
 f(\rho_{i}) = \rho_{i} \ctg\psi \quad ,
\end{equation}
where $\rho_{i}$ denotes the distance from the cone axis at which the $i$-th 
interface touches the substrate, and
\begin{equation}
\label{warbrz1}
 \left(1+\left(\dzup{f_{i}}{\rho}{}\right)^{2}\right)^{-1/2} 
   \left(1 + \ctg\psi \dzup{f_{i}}{\rho}{}\right) =
   (-1)^{i+1} \frac{1}{\sin\psi} 
   \frac{\sigma_{sv}-\sigma_{sl}}{\sab} - \frac{\eta_{slv}}{
   \sab \rho_{i}} \quad ,
\end{equation}

where all the derivatives are calculated at $\rho=\rho_{i}$.
\Eref{warbrz1} is equivalent to the statement that the 
contact angle $\theta_{i}$ of $i$-th interface (defined as the angle between 
this interface and the substrate measured across the liquid, see 
\fref{fsystem}) fulfills the following form of the modified Young equation 
for the cone
\cite{RW,BaN}
\begin{equation}
\cos\theta_{i} = \cos\theta_{0} + (-1)^{i+1}\frac{\eta_{slv}}{\sab \rho_{i}} 
\quad .
\end{equation}
$\theta_{0}$ denotes the Young angle, i.e. the contact angle of 
the liquid drop on a {\it flat} substrate and in the case of zero line 
tension 
\begin{equation}
 \cos\theta_{0} = \frac{\sigma_{sv}-\sigma_{sl}}{\sab} \quad .
\end{equation}
The solution of \eref{rownobj} which is smooth on the cone axis and 
satisfies the boundary condition \eref{defri} has the form 
\begin{equation}
\label{ogfi}
f_{i}(\rho) = \rho_{i} \ctg\psi + (-1)^{i} \sgn(\lambda) \left( 
\sqrt{\left(\frac{2\sab}{\lambda}\right)^2-\rho^{2}} -
  \sqrt{\left(\frac{2\sab}{\lambda}\right)^{2}-\rho_{i}^{2}}\right) .
\end{equation}
Thus each liquid-vapour interface forms a part of a sphere with radius 
$R = 2\sab/\lambda$. It can be determined from the fixed volume 
constraint \eref{warobj1} 
\begin{eqnarray}
\label{ogR}
\fl R = \left(\frac{3V}{\pi}\right)^{1/3} 
  \left(\sum_{i=1}^{n} \left(\cos^{2}(\theta_{i} + (-1)^{i+1}\psi)
  (-1)^{i+1} \frac{\cos\theta_{i}}{\sin\psi} + \right.\right. \nonumber\\
\left.\left. - 2(1-\sin(\theta_{i} + 
     (-1)^{i+1}\psi)) \right)\right)^{-1/3} \quad .
\end{eqnarray}
The location of the three-phase contact line of the $i$-th 
interface is given by 
\begin{equation}
\label{lokri}
 \rho_{i} = R \cos(\theta_{i} + (-1)^{i+1}\psi) \quad . 
\end{equation}

After introducing the dimensionless quantities 

\begin{equation}
 \bar{\cal F} = \frac{\cal F}{2 \pi (\frac{3V}{\pi})^{2/3} \sab}
   \quad ,
\end{equation}
\begin{equation}
 \bar{R} = \frac{R}{(\frac{3V}{\pi})^{1/3}} \quad ,
\end{equation}
\begin{equation}
\label{defetabar}
 \bar{\eta} = \frac{\eta_{slv}}{\sab
   (\frac{3V}{\pi})^{1/3}} \quad .
\end{equation}

the free energy of a given drop configuration can be expressed 
in the following form 
\begin{eqnarray}
\label{bFog}
\fl \bar{\cal F}_{n} = \bar{R}^{2} \sum_{i=1}^{n} \left(1 - \sin(\theta_{i} + 
(-1)^{i+1}\psi) \right. + \nonumber\\
\left. (-1)^{i} \frac{\cos\theta_{0}}{2\sin\psi} \cos^{2}(\theta_{i} +
(-1)^{i+1}\psi) + \frac{\bar{\eta}}{\bar{R}} \cos(\theta_{i} +
(-1)^{i+1}\psi) \right) \quad ,
\end{eqnarray}
where the subscript $n$ in $\bar{F}_{n}$ refers to the case with exactly $n$-
interfaces present in the system. The modified Young equation takes the form 
\begin{equation}
\label{rownnathi}
 \cos\theta_{i} - \cos\theta_{0} + \frac{\bar{\eta}}{\bar{R}}
  \frac{(-1)^{i+1} \sin\psi}{\cos(\theta_{i} + (-1)^{i+1}\psi)} = 0
  \quad .
\end{equation}
It can be checked that equations \eref{bFog} and \eref{rownnathi}
remain valid for all contact angles, so our 
primary assumptions that $\vartheta_{i}<\pi-\psi$ may be discarded. 

In the above formulae - in addition to the contact angles $\theta_{i}$ - 
only the parameters $\theta_{0}$, $\bar{\eta}$ and 
$\psi$ appear. Thus from now on this set of three variables  will be 
used to parameterize the equilibrium properties of the system. Note that 
the volume $V$ enters the rhs of \eref{bFog} only via the expression for 
$\bar\eta$ where it rescales the line tension. 

\section{Equilibrium drop configurations}

\subsection{Zero line tension}

\begin{figure}
\begin{center}
 \includegraphics{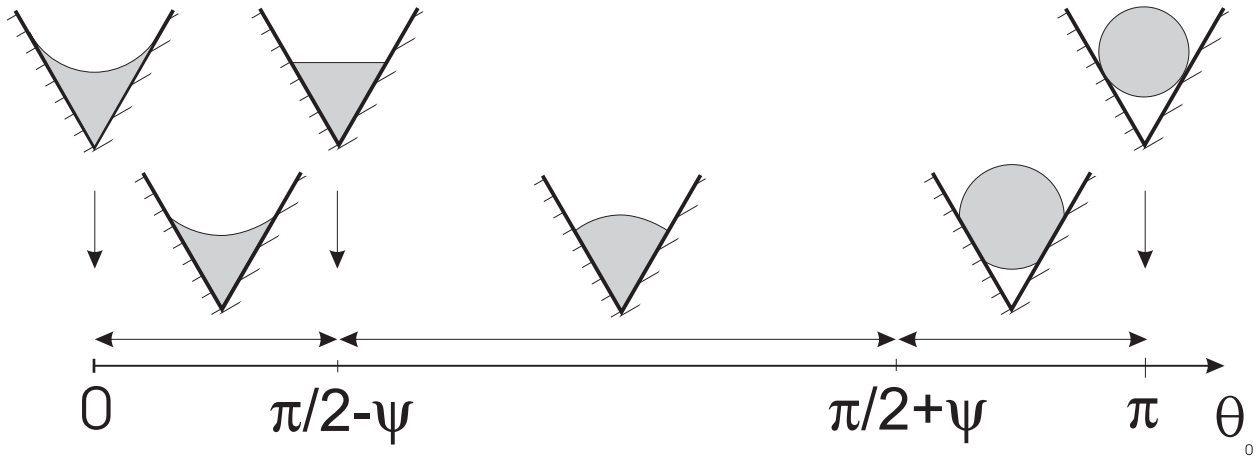}
\end{center}
\caption{The phase diagram in the case of zero line tension. The point 
$\theta_{0}=\psi+\pi/2$ corresponds to the continuous transition between the 
blob and the bridge configurations.\label{fdiafaz0}}
\end{figure}

In this subsection we put $\bar{\eta}=0$. In this case the solution of 
\eref{rownnathi} takes the form 
\begin{equation}
\label{rozwe0}
 \theta_{i} = \theta_{0} \quad 
\end{equation}
which means that only configurations with one and 
two interfaces are allowed. These configurations we will be called  
the blob and the bridge configuration, respectively \cite{CF, RTr}. The 
corresponding free energies take the following form  
\begin{equation}
\label{bF1e0}
 \bar{F}_{1} = 2^{-2/3} \left( (1 - \sin(\theta_{0}+\psi)) 
  - \frac{\cos\theta_{0}}{2\sin\psi} \cos^{2}(\theta_{0}+\psi) \right)^{1/3}
   \quad 
\end{equation}
(blob configuration) and 
\begin{eqnarray}
\label{bF2e0}
\fl \bar{F}_{2} =  2^{-2/3} \left( (1 - 
  \sin(\theta_{0}+\psi)) + (1 - \sin(\theta_{0}-\psi)) + \right. \nonumber\\
  \left. - \frac{\cos\theta_{0}}{2\sin\psi} (\cos^{2}(\theta_{0}+\psi) 
  - \cos^{2}(\theta_{0} - \psi)) \right)^{1/3} \quad 
\end{eqnarray}
(bridge configuration). 
Comparing the above free energies leads to the conclusion 
that for $\theta_{0}\leq\pi/2+\psi$ the equilibrium configuration has the 
form of a blob while  for $\theta_{0}>\pi/2+\psi$ it takes the form of a  
bridge.  
According to \eref{lokri} the transition between the blob and the bridge 
configurations corresponds to the situation in which the bottom interface of 
the bridge reaches the vertex of the cone. This transition is continuous. 
This results is similar to the one for adsorption in a wedge \cite{RTr}. 
\Fref{fdiafaz0} shows the relevant phase diagram. \\ 

\subsection{Non-zero line tension}

For $\eta\neq 0$ the analysis of \eref{rownnathi} shows that - similarly 
to the case of zero line tension - only configurations 
with one and two interfaces are admitted. This case, contrary to 
the previous one can be treated only numerically. One method of finding 
the equilibrium values of the contact angles is to solve numerically
\eref{rownnathi} and then to calculate the corresponding values of 
the free energy in order to find the equilibrium solution.  
However, it turns out that the direct numerical minimization of the 
constrained free energy is more efficient procedure in  
determining the equilibrium configurations and we follow this method 
\cite{sumsl}. 

The equilibrium configuration of the drop - for given set of $\theta_{0}, 
\bar\eta$, and $\psi$ values - corresponds to the global minimum of 
the free energy. In practice one has to find all the minima of \eref{bFog} 
corresponding to the blob ($n=1$) and the bridge ($n=2$) configurations 
including also the minima localized on the borders of contact angle domain, 
i.e. $\theta_{1}$ or $\theta_{2}$ equal to $0$ or $\pi$, and then to compare 
the values of the corresponding free energy. \Tref{konf} introduces the 
terminology used in describing the phase diagram. 

\begin{figure}
\begin{center}
 \includegraphics{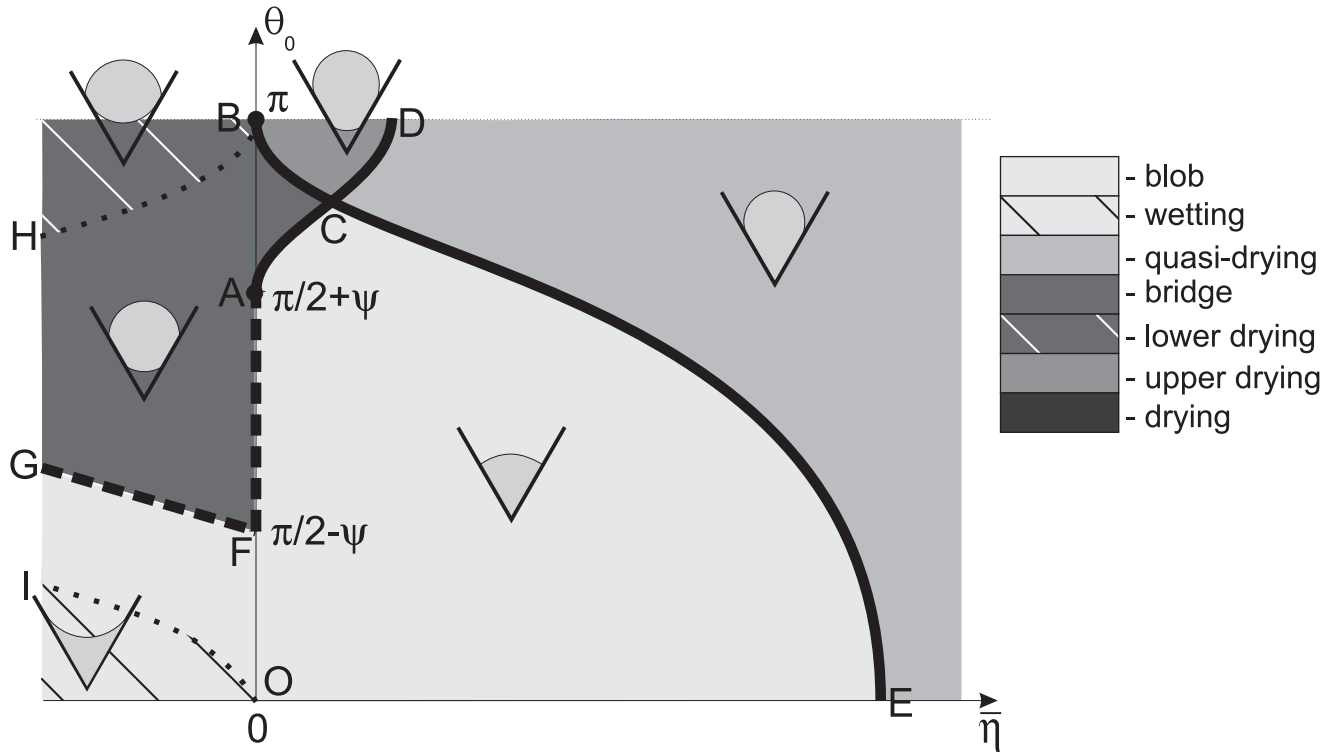}
\end{center}
\caption{The generic phase diagram. Different shades of gray correspond 
to different phases which are schematically depicted in the insets. The 
continuous lines B-E and A-D correspond to the first-order phase 
transitions; the broken line  A-F-G corresponds to the continuous 
transition. Point A denotes the tricritical point. The dotted line IO forms 
the border of the region in which the contact angle $\theta_{1}=0$; the 
dotted line HB forms the border of the region in which the contact angle 
$\theta_{2}=\pi$.\label{fdiafaz}}
\end{figure}

\begin{figure}
\begin{center}
 \includegraphics{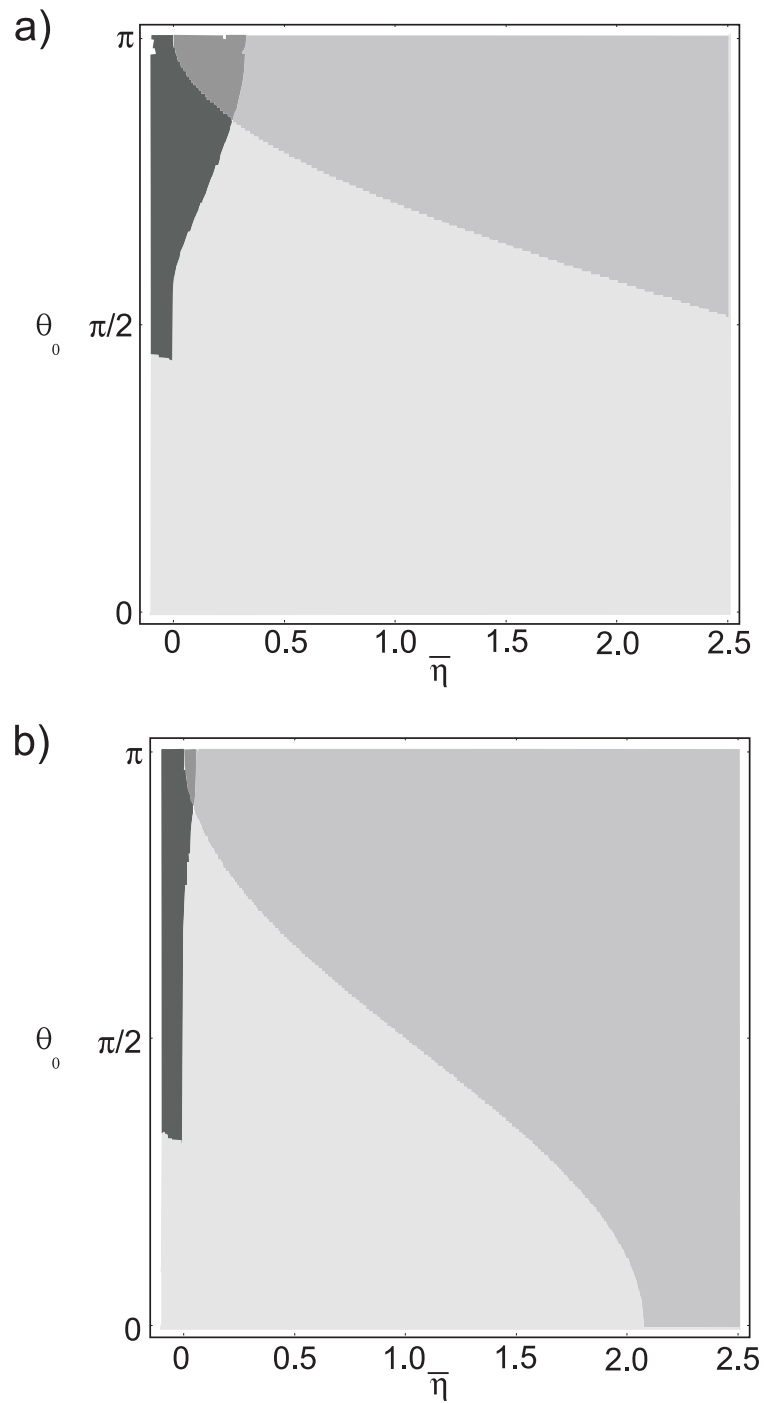}
\end{center}
\caption{The phase diagrams for two particular choices of the cone opening 
angle: a) $\psi=\pi/16$, b) $\psi=3\pi/16$. The shades of gray 
correspond to those used in \fref{fdiafaz}. Note that the regions 
corresponding to the wetting and lower drying configurations are not marked 
on this figure.
\label{fdiafaza}}
\end{figure}

\begin{table}
\caption{Configurations of liquid in a cone. \label{konf}}
\begin{indented}
\item[]
\begin{tabular}{lccc} 
\br
 configuration & number of interfaces & $\theta_{1}$ & $\theta_{2}$ \\ 
\mr
blob & 1 & $\neq 0$, $\neq \pi$ & - \\ wetting & 1 & $=0$ & - \\ 
quasi-drying & 1 & $=\pi$ & - \\ 
bridge & 2 & $\neq 0$, $\neq \pi$ & $\neq 0$, $\neq \pi$ \\ 
lower drying & 2 & $\neq 0$, $\neq \pi$ & $=\pi$ \\ 
upper drying & 2 & $=\pi$ & $\neq 0$, $\neq \pi$ \\ 
drying & 2 & $=\pi$ & $=\pi$ \\ 
\br
\end{tabular}  
\end{indented} 
\end{table}
\Fref{fdiafaz} shows schematically the generic phase diagram; \fref{fdiafaza} 
presents the actual phase diagrams corresponding to particular values of the 
cone opening angle. \\

Before discussing the phase diagram it is worthwhile to recall that for given 
value of the cone opening angle $\psi$  
the phase diagram is constructed in function of the Young angle $\theta_{0}$, 
eq.(8)  and the \it dimensionless \rm line tension $\bar\eta$, Eq.(12c) which 
is inversely proportional to the cubic root of the liquid volume $V$. This 
means that substantial influence of the line tension can be observed only for 
drops of sufficiently small volume. If one takes the typical value of line 
tension to be $10^{-8} N$ \cite{LT}, the typical value of liquid-vapour 
surface tension to be $10^{-2} N/m$ \cite{ST}, and the size of the liquid 
drop to be of micrometer size then the dimensionless line tension is of order 
unity. For much bigger drops the effects of line tension will be 
negligible. \\ 

The rich structure of the phase diagram results from the competition between 
the surface tension and the line tension. Depending on the values of 
parameters $\theta_{0}$ and $\bar\eta$ the system can be found in one of 
four different phases. They are distinguished by the number of interfaces 
present in the system (one or two) and the value of the contact angle 
$\theta_{1}$ (corresponding to the upper interface) being equal to $\pi$ 
or being different from $\pi$, see also Table 1. These four phases are 
denoted on \fref{fdiafaz} with the 
help of different shades of gray and are separated from each other by thick 
continuous or thick broken lines depending on the order of the transition.
To make the phase diagram more transparent each phase is schematically
marked by the corresponding liquid configuration. 
Additionally, within two of the above phases one can distinguish 
quantitatively different regions. The lower left part of the phase diagram 
contains region corresponding to configuration with only one interface and 
zero contact angle; we call it the wetting configuration. The dotted line 
I-O marks the boundary of this region; upon crossing this line the contact 
angle ceases to be equal to $0$. The upper left part of the phase diagram 
corresponds to configuration with two interfaces and the contact angle 
$\theta_{2}$ corresponding to the lower interface being equal to $\pi$. 
Again the dotted line H-B marks the boundary of this region; upon crossing 
it the contact angle $\theta_{2}$ ceases to be equal to $\pi$. \\  
The line B-E is the first-order transition line between configurations 
distinguished by the value of the contact angle $\theta_{1}$ being equal 
to $\pi$ and different from $\pi$. Note that the segment C-E 
corresponds to the first-order transition between phases in which only 
one interface is present while the segment B-C corresponds to the first-order 
transition between the phases in which two interfaces are present; upon
crossing this segment discontinuity of $\theta_{2}$ is observed. It is 
worthwhile to note that the line B-E corresponds - for the cone geometry 
considered in this paper - to the phase transition found by Widom \cite{W12} 
when analyzing a liquid drop sessile on a flat substrate. 
The difference between the phases separated by the line B-E becomes 
smaller and smaller upon approaching point B and finally disappears at this 
point. To examine the shape of the B-E line in the neighborhood of this point 
we look for approximate analytical solution of \eref{rownnathi}. For  
$\bar{\eta} \rightarrow 0$ this set of equations can be solved 
perturbatively and one finds that
\begin{equation}
(\pi - \theta_{0}) \sim \bar{\eta}^{1/2} \quad .
\end{equation} 
The G-F-A-D line is the line of transitions between configurations with one 
and with two interfaces, respectively. Along the segment G-F-A this 
transition is continuous while along the segment A-D it is first-order. It 
corresponds to discontinuity of the value of the contact angle $\theta_{2}$ 
upon crossing the segment A-C one additionally observes the 
discontinuity  of $\theta_{1}$. 
Thus point A ($\eta=0$, $\theta_{0}=\pi/2 + \psi$), which played a 
distinguished
role on \fref{fdiafaz0}, tends out to be the tricritical point in the enlarged
space of system parameters. 
Analyzing the shape of the A-D line in the vicinity of the tricritical point 
leads again to the conclusion that
\begin{equation} 
\left(\theta_{0} - \left(\frac{\pi}{2} + \psi \right) \right) \sim 
\bar{\eta}^{1/2} \quad .
\end{equation}
One also observes that upon approaching the tricritical point the 
discontinuity of $\theta_{2}$ decreases to zero as the square root of 
$\bar{\eta}$. \\ 
As far as the dependence on the cone opening angle $\psi$ is considered one 
observes the decrease of the region BACD on \fref{fdiafaz} upon increasing 
$\psi$ towards $\pi/2$; in this limit points A and D tend towards point B.
This area corresponds to situations in which positive line 
tension and the values of $\theta_{0}$ close to $\pi$ stabilize the 
configurations with two interfaces. Finally, for $\psi=\pi/2$ one recovers
the phase diagram discussed by Widom \cite{W12}. \\

\section{Conclusions}

In these paper we have investigated the configurations of a non-volatile 
liquid drop placed in a cone. Our macroscopic analysis based on the 
constrained free energy 
leads to the phase diagram parameterized by the Young angle $\theta_{0}$ and 
the dimensionless line tension $\bar\eta$. The interesting feature of the 
phase diagram is the existence of lines of the first- and the second-order 
transitions corresponding either to discontinuous or continuous changes of 
the contact angles. One also finds the tricritical point and the parabolic 
shape of the transition line in its vicinity. Another property of the system 
is that the negative line tension stabilizes the so-called 
wetting configuration into the range of non-zero values of the Young angle. 
This wetting configuration is characterized by the presence of a single 
interface and zero contact angle. 

One should keep in mind that the present analysis is restricted to
interfacial configurations which are cylindrically symmetric and smooth at
the cone axis. In the case of negative line tension with large absolute 
values 
one might expect still a different class of equilibrium configuration which 
consist of separate "rings" extending along the cone.
Analogous situations appear also in the analysis of a sessile drop on flat 
substrate; this will be the subject of our future analysis. \\

\section*{Acknowledgment}
The authors gratefully acknowledge the discussions with prof. Hans W. Diehl 
and  prof. Herbert Wagner, and the support by the Foundation for 
German-Polish Collaboration under Grant. No. 3269/97/LN.

\end{document}